\documentclass{emulateapj}

\shorttitle{The Environment of a Large Ly$\alpha$ Nebula}
\shortauthors{Prescott et al.}

\newcommand{\lya}{Ly$\alpha$}
\newcommand{\lab}{LAB}
\newcommand{\lae}{LAE}
\newcommand{\labs}{LABs}
\newcommand{\laes}{LAEs}
\newcommand{\lbg}{LBG}

\newcommand{\lbgs}{LBGs}
\newcommand{\smgs}{SMGs}

\newcommand{\bw}{$B_{W}$}
\newcommand{\ib}{$IA445$}

\newcommand{\cgs}{erg s$^{-1}$ cm$^{-2}$}

\begin{document}

\title{The Overdense Environment of a Large L\lowercase{$y$}$\alpha$ Nebula at \lowercase{$z$}$\approx$2.7\altaffilmark{1,2}}
\author{Moire K. M. Prescott\altaffilmark{3}, Nobunari Kashikawa\altaffilmark{4}, Arjun Dey\altaffilmark{5}, Yuichi Matsuda\altaffilmark{6}}
\altaffiltext{1}{Based on data collected at Subaru Telescope, 
which is operated by the National Astronomical Observatory of Japan.}
\altaffiltext{2}{Based in part on observations obtained at Kitt Peak National Observatory, a division of the National Optical 
Astronomy Observatories, which is operated by the Association of Universities for Research in Astronomy, Inc. 
under cooperative agreement with the National Science Foundation.}
\altaffiltext{3}{Steward Observatory, University of Arizona, 933 North Cherry Avenue, Tucson, AZ 85721; mprescott@as.arizona.edu}
\altaffiltext{4}{National Astronomical Observatory, Mitaka, Tokyo 181-8588, Japan; kashik@zone.mtk.nao.ac.jp}
\altaffiltext{5}{National Optical Astronomy Observatory, 950 North Cherry Avenue, Tucson, AZ 85719; dey@noao.edu}
\altaffiltext{6}{Department of Astronomy, Kyoto University, Sakyo-ku, Kyoto 606-8502, Japan; matsdayi@kusastro.kyoto-u.ac.jp}

\begin{abstract}
Large nebulae ($\gtrsim$50~kpc) emitting strongly in \lya\ (also known as Ly$\alpha$ ``blobs'') 
are likely signposts of ongoing massive galaxy 
formation.  The relative rarity of these sources and their discovery in 
well-studied galaxy overdensities suggest that they may be associated with regions of high galaxy density.  
One of the largest \lya\ nebulae, discovered at a redshift of $z\approx$~2.7 via its strong mid-infrared emission, 
provides an unbiased test of this association.  We have carried out a deep intermediate-band imaging survey 
for \lya-emitting galaxies (\laes) within a 30\arcmin$\times$26\arcmin\ 
field of view around this \lya\ nebula.  This is the first study 
of the environment of a \lya\ nebula found without a priori knowledge of its surroundings.  We find that 
the nebula is located in an overdense region, at least 20$\times$50~h$_{70}^{-1}$~comoving Mpc in size, 
showing a factor of $\sim$3 \lae\ number density enhancement relative to the edge of the field.  
Given the predicted number of such overdensities, we rule out the 
possibility of a chance coincidence at the $\lesssim$1\% level.  
This study, in conjunction with previous work, provides strong confirmation of the association between the 
largest \lya\ nebulae and overdense regions of the Universe.
\end{abstract}


\section{Introduction}

Studies of massive galaxy populations show that the most massive galaxies are in place 
and in possession of the majority of their stellar mass by $z\sim1-2$ 
\citep{mcc04,van04,dad05,bun05,bro07}.  
Thus, while dark matter halos in a $\Lambda$CDM cosmology build up 
hierarchically with the gradual accretion and merging of smaller halos, 
the most massive galaxies within that context likely have more 
dramatic origins.  The details of this process are uncertain, and ideally we would like 
to study sites of ongoing massive galaxy formation.  \lya\ nebulae or \lya\ ``blobs'' 
(LABs) -- large ($\gtrsim$50~kpc) clouds of gas emitting strongly 
in \lya\ ($\sim10^{44}$ erg s$^{-1}$) -- provide such an 
opportunity.  The large \lya\ equivalent widths and the 
association with galaxy populations such as Lyman 
break galaxies (\lbgs) and submillimeter galaxies (\smgs) 
strongly suggest that \labs\ are sites of ongoing 
galaxy formation.  They have been found in small numbers (only 16 $\gtrsim$50~kpc \labs\ are known) 
around $z\sim$~2-3, a key epoch of black hole and galaxy growth 
(\citealt{stei00}, hereafter S00; \citealt{fra01,pal04,mat04,dey05,nil06,smi07,gre07}).  
\labs\ span a range in size and surface brightness from the 
scales of \lya-emitting galaxies (\laes) to the largest \labs\ 
known \citep[$\sim$150~kpc, $\sim$27 mag arcsec$^{-2}$; e.g.,][]{mat04}.  
\labs\ are similar to the large \lya\ halos observed in the overdensities around 
higher redshift radio galaxies \citep[e.g.,][]{ven07,ove08} but are found 
in radio quiet environments.  
The dominant power source in \labs\ is difficult to determine; 
among the limited sample of large \labs, there is evidence 
for embedded AGN, starburst-driven superwinds, gravitational cooling 
radiation, and spatially-extended star formation, all of which may 
play a role in powering the \lya\ emission (Prescott et al. 2008, in preparation; 
S00; \citealt{chap04, nil06, tani00, mat04, dey05, mat07}).

Most known \labs, including two of the largest cases, have been discovered via narrow-band 
surveys, often by targeting known galaxy overdensities 
(S00; \citealt{fra01, pal04, mat04}).  
Follow-up narrow-band imaging of the SSA22 region surrounding the S00 LABs 
revealed fainter \labs\ associated with the same overdensity traced by the 
\laes\ \citep{mat04}.  This suggests that \labs\ may be confined to overdense regions, 
as would be expected for sites of massive galaxy formation.  
A blank field survey by \citet{sai06} supported this claim, finding 
no \labs\ with sizes greater than $\sim$30~kpc and an 
order of magnitude lower number density of \labs\ relative to 
that found within the SSA22 galaxy overdensity.  
However, the association between \labs\ and galaxy overdensities 
may be misleading, as a truely systematic 
wide-area search has yet to be completed.  A thorough environmental 
study has only been done for the S00 \labs, which were 
found by targeting a known galaxy overdensity.  

In contrast, one of the largest \labs\ uncovered 
recently was found by entirely different means.  
While conducting a study of  mid-infrared (24$\mu$m)
sources detected by the {\it Spitzer Space Telescope}, 
\citet{dey05} discovered a \lab\ at $z=2.656$ within 
the NOAO Deep Wide-Field Survey Bo\"otes field \citep[NDWFS\footnote{This research 
draws upon data provided by Dr. Buell Jannuzi and Dr. Arjun Dey as distributed by the 
NOAO Science Archive.}; ][]{jan99}.  
Follow-up observations revealed complexity typical 
of the \lab\ class: the region hosts a buried AGN, many 
young galaxies, an \lbg, and diffuse He\textsc{II} 
and continuum emission suggestive of spatially-extended star formation 
(Prescott et al. 2008, in preparation).  One of the largest and most luminous \labs\ known, 
and found without any a priori knowledge of its surroundings, 
this source represents a unique opportunity to 
perform an unbiased, complementary test of the association between \labs\ 
and overdense regions.  

In this work, we present the first results from an ongoing deep intermediate-band \lya\ imaging 
survey of the environment of the \citet{dey05} \lab\ (hereafter, LABd05) and report on the spatial distribution of the
\laes\ in the immediate vicinity.  
A detailed analysis of the multiwavelength properties of LABd05 and 
the properties and clustering of the LAEs in the region will be presented by 
Prescott et al. (2008, in preparation).  We assume the standard $\Lambda$CDM cosmology 
($\Omega_{M}$=0.3, $\Omega_{\Lambda}$=0.7, $h$=0.7); 
the angular scale at $z=2.656$ is 7.96 kpc/\arcsec.  All magnitudes are in the AB system.

\section{Observations \& Reductions}

We obtained deep imaging of the field around LABd05 using the Subaru telescope and 
the SuprimeCam 
imager \citep{miya02} on U.T. 2007 May 10-14 and June 17.  
The survey covers 0.22~deg$^{2}$ in an intermediate-band 
filter, IA445 ($\lambda_{c}\approx$4458\AA, 
$\Delta\lambda_{FWHM}\approx$201\AA), centered on the \lya\ 
line at the redshift of the nebula; this corresponds to a comoving 
volume of 4.27$\times$10$^{5}$ h$_{70}^{-3}$ Mpc$^{3}$ 
(52$\times$45$\times$180~h$_{70}^{-1}$~Mpc).  
Conditions during the May observations were variable 
(clouds, variable seeing 0.7\arcsec-1.2\arcsec) and good in June 
(clear with 0.7\arcsec\ seeing).  We obtained 
a total of 3 hours of observations.  

We reduced the data using the SDFRED software \citep{yagi02,ouchi04}.  
The data were overscan-subtracted and corrected for geometric and atmospheric 
distortions.  We generated the sky flat using object frames that 
were free of bright stars in combination with other images 
taken in the same intermediate-band filter (Y. Taniguchi, private communication).  
Small portions of the SuprimeCam field of view are vignetted by the autoguider probe; the 
affected areas were masked, as were bad columns, bright star ghosts, 
and satellite trails.  
Images were aligned and scaled using common stars and then combined 
using a clipped mean algorithm, which successfully removed cosmic rays.  
Of the SuprimeCam field-of-view (0.26~deg$^{2}$), 73\% was 
usable (the remainder being of lower signal-to-noise along 
the edge of the field or in the vicinity of bright stars).  

The limiting magnitude of the stacked image is 28.3 AB mag (1$\sigma$, 2\arcsec\ diameter aperture), 
calculated using 10,000 random apertures.  An approximate magnitude zeropoint was 
calculated from observations of the standard stars BD+25d4655 and Feige 34.  
For the NDWFS broad-band imaging, the limiting magnitudes are 
\bw$_{limit}$=27.9 mag, $R_{limit}$=27.1 mag, $I_{limit}$=26.0 mag 
(1$\sigma$, 2\arcsec\ diameter aperture).  

A portion of the field was observed 
in a custom $U$-band filter ($U_{d}$; $\lambda\approx$3590\AA, $\Delta\lambda_{FWHM}\approx$116\AA) 
using the Mayall 4m Telescope over 6 nights (U.T. 2007 June 8-13).  
These data will be described elsewhere (Prescott et al. 2008, in preparation) 
but here provide a useful check on interlopers in our $z\approx2.7$ \lae\ sample.  

\section{Candidate Selection}
  
\begin{figure}
\plotone{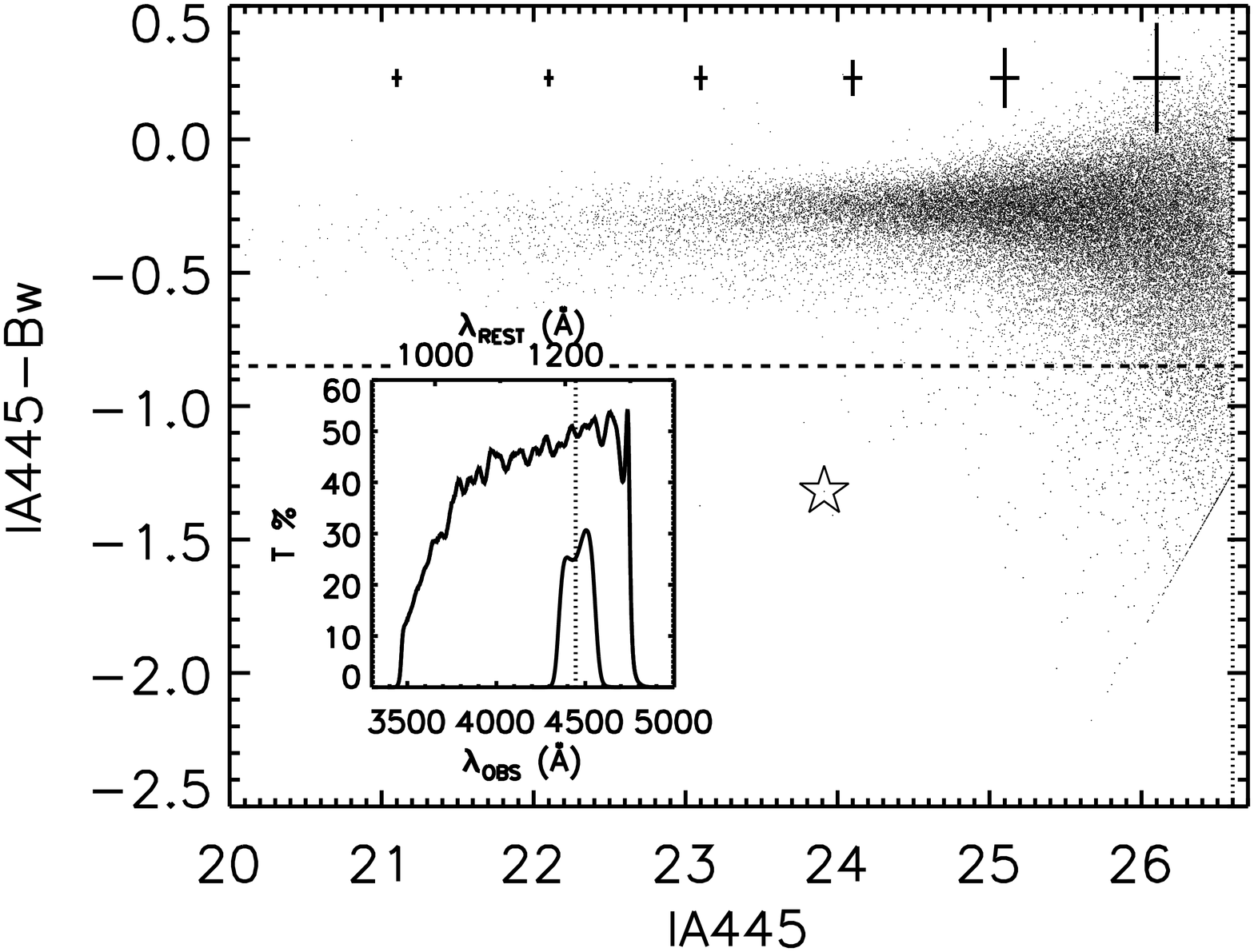}
\caption{\ib$-$\bw\ color-magnitude diagram for sources detected in the \ib\ image (dots).  
Initial candidates were selected to meet the color cut \ib$-$\bw$\leq-$0.85 (dashed line, 
which corresponds to an observed equivalent width of $W_{obs}\geq$148\AA) 
and to be brighter than the 5$\sigma$ limiting magnitude in \ib\ band (\ib$_{5\sigma,limit}$=26.6; dotted line).  
LABd05 is shown as a star, and non-detections in the \bw\ band are set to the 
\bw\ limiting magnitude.  Typical error bars are shown.  
The inset shows the effective transmission curves for the \bw\ and \ib\ 
filters and the wavelength of Ly$\alpha$ at the redshift of the system (dotted line).}
\label{fig:ibcolor}
\end{figure}

We used Source Extractor \citep{ber96} to select a sample of $\approx$38,600 sources 
detected in the \ib\ band down to the 5$\sigma$ limit of \ib=26.5 mag 
(2\arcsec\ diameter aperture; $L_{IA445}$($z=2.656$)=1.5$\times$10$^{42}$ erg s$^{-1}$) with the 
following search parameters: at least 5 contiguous pixels, a threshold of 2$\sigma$ per pixel, 
and a Gaussian filter matched to the seeing (FWHM$\approx$0.8 arcsec).  We measured matched 
aperture photometry using 2\arcsec\ diameter apertures from the \ib, \bw, $R$, and $I$ 
imaging datasets, which were registered and resampled to match the \bw\ astrometry 
and pixel scale (0.258 arcsec/pix).  
Aperture corrections were neglected ($\approx$1.08 for an unresolved source).  
Line-emitting sources are strongly detected in \ib\ relative to \bw, 
i.e., they have large negative \ib$-$\bw\ colors relative to the normal galaxy locus 
(see Figure~\ref{fig:ibcolor}).  We removed bright stars (\ib$\leq25.0$) 
using the CLASS\_STAR parameter in Source Extractor ($>$0.91) 
and employed a cut of \ib$-$\bw$\leq-$0.85~mag 
yielding 1500 candidates.  Shifting the \ib$-$\bw\ cut by $\pm$0.2~mag causes no significant 
change to the main results presented in Section 4.  For a $F_{\nu}\propto\nu^{0}$ continuum source, 
this corresponds to an observed equivalent width (EW) cut of $W_{obs}\geq$148~\AA.  
\laes\ are known to be young, with estimated ages of 4-200~Myr \citep{fink07,lai07,gaw07}.  
For the case of a young galaxy \citep[25~Myr old simple stellar population, solar metallicity, 
Chabrier IMF;][]{bc03,tre04} at $z=2.7$ with standard intergalactic absorption \citep{mad95}, 
this is equivalent to a rest-frame EW cut of $W_{rest}\geq$50\AA.

\begin{figure}
\plotone{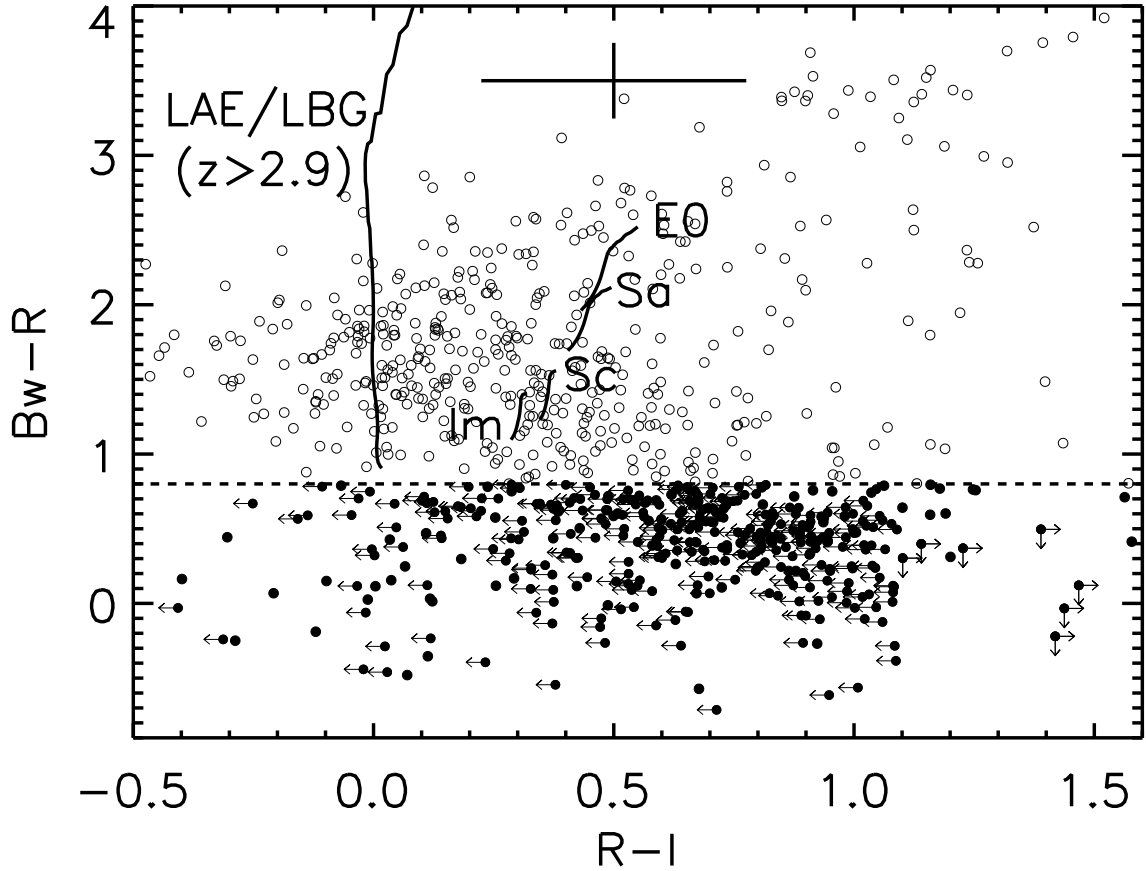}
\caption{\bw$-R$ vs. $R-I$ color-color diagram of the 1500 line-emitting candidates 
(open and filled circles).  A typical error bar is shown.  
Expected color tracks for low redshift galaxy templates are shown for 0.1$<z<$0.3 \citep{lei96}; 
the color track for high redshift \lae/\lbg\ 
contaminants is based on a young galaxy template at 2.9$<z<$4.0 
\citep[25~Myr old simple stellar population, solar metallicity, Chabrier IMF;][]{bc03} 
with standard intergalactic absorption \citep{mad95}.  
To remove both low and high redshift interlopers, candidates were required to 
have \bw$-R\leq$0.8.  The final sample of 785 \lae\ candidates is shown (filled circles).}
\label{fig:colorcolor}
\end{figure}

We expect our \lae\ candidate sample to be contaminated 
by high EW, low-redshift [OII]$\lambda\lambda$3727,3729-emitting galaxies at $z\approx0.2$.  
Since the \ib\ bandpass lies on the red side of the \bw\ filter (Figure~\ref{fig:ibcolor}, 
see inset), we also expect contamination from higher redshift galaxies ($z\gtrsim2.9-4.0$) 
for which the Lyman limit has entered the \bw\ filter, thus depressing the \bw\ flux 
relative to the \ib\ flux.  
Using the publicly available NDWFS imaging \citep{jan99}, 
we employ a cut of \bw$-R\leq$0.8 to remove both contaminant populations 
(see Figure~\ref{fig:colorcolor}).  
The final sample of 785 \lae\ candidates 
corresponds to a mean \lae\ surface density of $\approx$4200 deg$^{-2}$.  
The properties and sizes of the \lae\ sample will be discussed 
in an upcoming paper (Prescott et al. 2008, in preparation).  
There are no other large ($\gtrsim$50~kpc) \labs\ 
in the vicinity of LABd05.

Sources at these redshifts should show very little flux in the $U_{d}$-band, which straddles 
the Lyman limit ($\lambda_{rest}\approx$970-1000\AA\ at $z\approx2.656$).  
Using the 25~Myr old model above to represent the typical \lae\ continuum shape, 
the $z\approx3$ galaxy luminosity function from \citet{red07}, and 
standard intergalactic absorption \citep{mad95}, 
we predict that $\sim$1-2\% of the \lae\ sample should have $U_{d}$ 
detected at the 5$\sigma$ level ($U_{d}$=25.3).  
Only 1 \lae\ candidate ($<$1\%) is detected with $U_{d}\leq$25.3, 
giving us confidence that we have selected a clean sample of \laes.  
More sophisticated interloper rejection would require deeper $U_{d}$ imaging 
over the entire field.  

\section{Results and Discussion}
\begin{figure}
\plotone{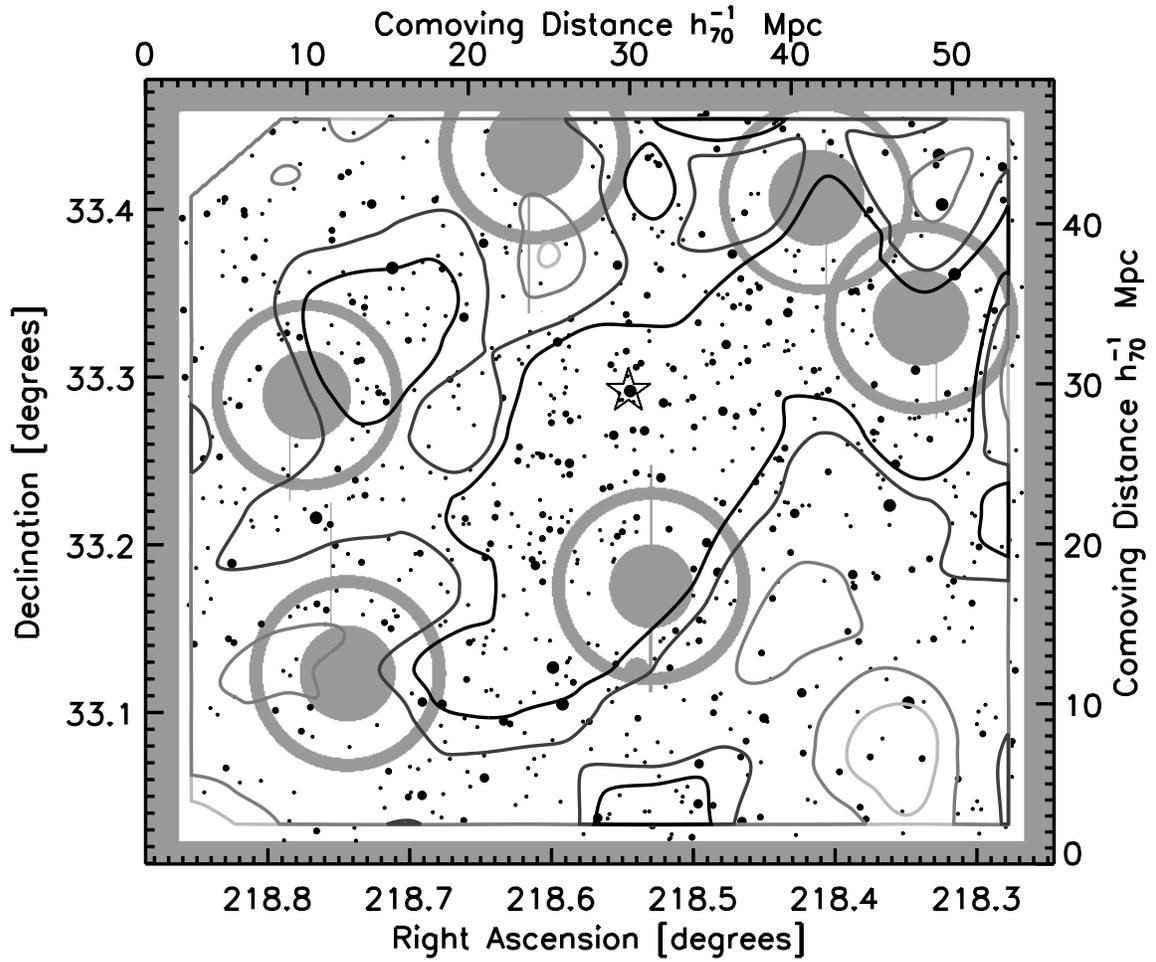}
\caption{Spatial distribution of \lae\ candidates (filled circles).  
After smoothing on 3.2\arcmin$\times$2.8\arcmin\ scales, isodensity contours were laid down 
at 0.7, 1.0, 1.8, and 2.3 times the average \lae\ density at the edge of the field.  
The four circle sizes represent four bins in \ib\ luminosity: 
$L_{IA445} \leq 2.5\times10^{42}$, $2.5\times10^{42}<L_{IA445}\leq5\times10^{42}$, 
$5.0\times10^{42}<L_{IA445}\leq1.0\times10^{43}$, $L_{IA445}>1.0\times10^{43}$.  
The position of LABd05 is indicated with a star.  
Regions masked along the edge and due to bright stars are shown in grey.  There is evidence 
for an extended structure stretching from SE to NW and a $\sim$10~h$_{70}^{-1}$~Mpc 
void at [218.36,33.07].}
\label{fig:spatial}
\end{figure}

\begin{figure}
\plotone{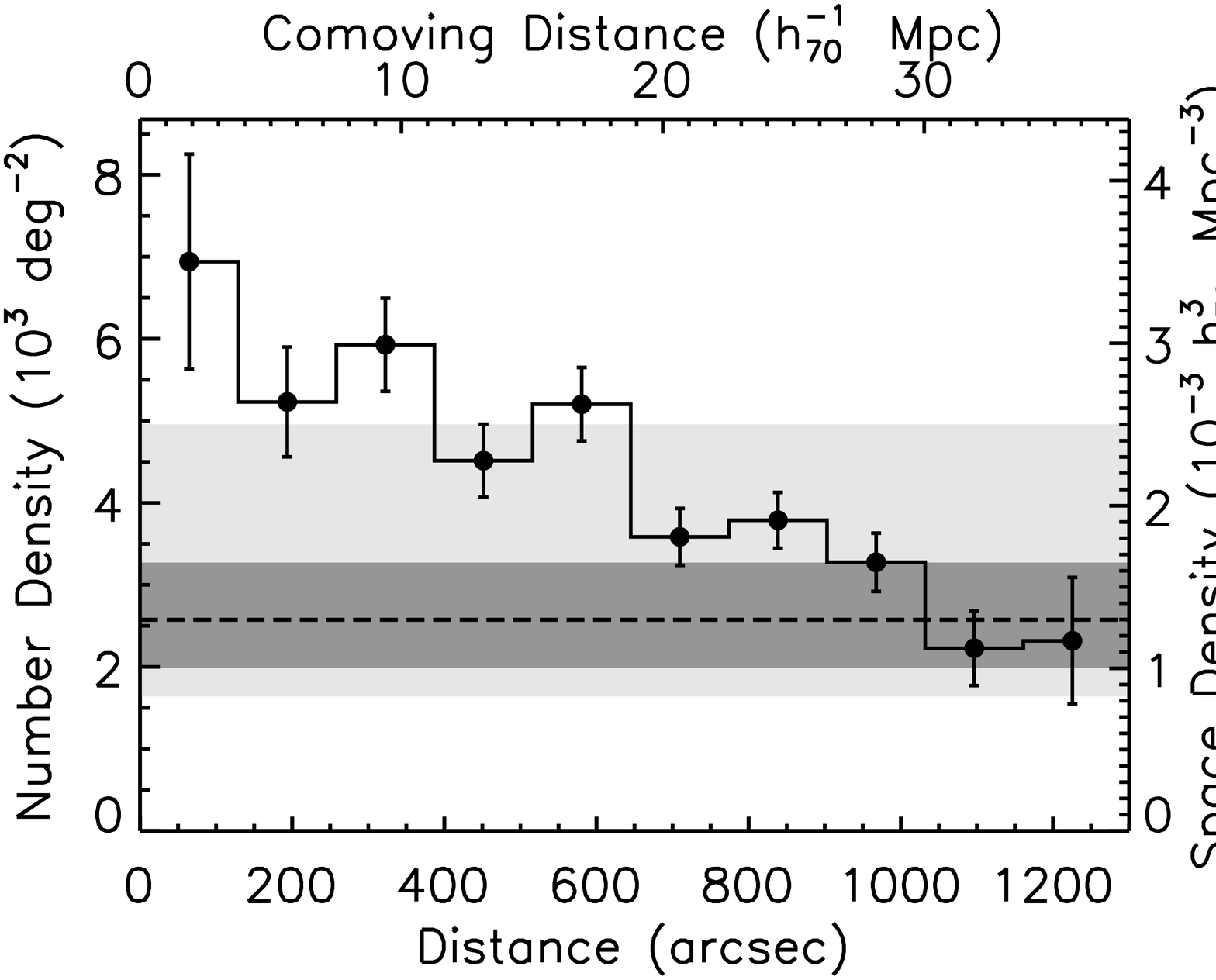}
\caption{The radial density profile measured in 
circular annuli centered on LABd05.  
The \lab\ lies in an overdense region that is a 
factor of $\sim$3 times the density at the edge of the field of view.  
The dashed line represent the predicted number density 
at $z\approx3$ if we assume a uniform redshift distribution 
and no evolution \citep{gron07,ouc07}.  
The shaded areas correspond to the range covered by 1$\sigma$ error bars; 
all three parameters ($\alpha$,$L^{*}$,$\phi^{*}$) from \citet{gron07} were 
allowed to vary in turn (light shading), while in the comparison to 
\citet{ouc07} the faint-end slope $\alpha$ is fixed at $-$1.5 (dark shading).  
The predicted \lae\ number density is consistent with what 
we measured at the edge of the field.  Since the 
redshift distribution of our sample can only be narrower than assumed, 
the overdensity measured within the structure is likely a lower limit.}
\label{fig:density}
\end{figure}

Figure~\ref{fig:spatial} shows the spatial distribution of 
\laes\ in the vicinity of LABd05.  We find an overabundance of \laes\ 
in the immediate vicinity of the \lab\ and clear evidence of an elongated overdense 
structure, with what appears to be a $\sim$10~h$_{70}^{-1}$~Mpc underdensity at [218.36,33.07].  
In Figure~\ref{fig:density}, we plot the surface density of \laes\ versus distance from LABd05.  
This shows a peak overdensity of a factor of 3.0 relative to the edge of the field, 
or a factor of 2.6 when averaged over the central 10~h$_{70}^{-1}$~Mpc radius.  
For comparison, the dashed line shows the expected \lae\ surface density estimated from 
the $z\approx3.1$ luminosity functions of \citet{gron07} and \citet{ouc07} assuming a uniform 
redshift distribution across our bandpass and no evolution.  
The overdensity spans at least $\approx$17$\times$47~h$_{70}^{-1}$~Mpc (comoving).  
A study of the \lae\ population around the 
two S00 \labs, which are of comparable size and luminosity 
to LABd05, also found a factor of $\sim$3 overdensity relative 
to the field that was $\sim$60~Mpc across \citep{hay04}.  
Thus, although it was found without any a priori knowledge of its surroundings, 
LABd05 appears to reside in a similarly overdense environment.  

Without follow-up spectroscopy, we cannot know the true redshift distribution of the sample 
within the intermediate-band filter.  Assuming a uniform distribution across the filter 
($\Delta z \approx$~0.17, 180~h$_{70}^{-1}$~Mpc, 
comoving, along the line-of-sight), our survey yields 
a mean number density of $\rho\approx$2.1$\times$10$^{-3}$~h$_{70}^{-3}$~Mpc$^{-3}$ for \laes\ with 
$L_{IA445}\gtrsim$1.5$\times$10$^{42}$ \cgs.  The inner contour in Figure~\ref{fig:spatial} 
corresponds to $\rho\approx$2.8$\times$10$^{-3}$~h$_{70}^{-3}$~Mpc$^{-3}$ while 
at the edge of the field $\rho\approx$1.2$\times$10$^{-3}$~h$_{70}^{-3}$~Mpc$^{-3}$.  
Given the presence of an overdensity, we expect the true redshift distribution to 
be significantly narrower.  If we assume the overdensity resembles that hosting the 
S00 \labs, which extends 40~h$_{70}^{-1}$~~Mpc comoving (22\% of the filter width) along the 
line-of-sight \citep{mat05}, and that it is centered within the \ib\ filter, the inner 
contour corresponds to 
a number density of $\rho\approx$12$\times$10$^{-3}$~h$_{70}^{-3}$~Mpc$^{-3}$.  
Alternately, if the structure is cylindrical in shape with a radius of 10~h$_{70}^{-1}$~Mpc, 
the corresponding number density at the inner contour is $\rho\approx$25$\times$10$^{-3}$~h$_{70}^{-3}$~Mpc$^{-3}$.    

We estimate the expected frequency of such overdense structures 
using the galaxy catalog of \citet{bow06}, which is based on 
the Millennium Simulation \citep{spri05}.  
We approximate our observational set-up by sampling the simulation volume randomly 
with cylinders that are 10~h$_{70}^{-1}$~Mpc in radius (the size of the 
overdensity peak) and 180~h$_{70}^{-1}$~Mpc in depth (the full span of the filter).  
We assumed a Ly$\alpha$/H$\alpha$ ratio of 10 \citep{ost89} and 
scaled from the predicted H$\alpha$ luminosity to select model \lae\ galaxies
down to the \lya\ limit of our observations (under the assumption 
that \lya\ dominates the measured \ib\ flux) at a redshift of $z\approx2.7$.  
The number density of model \laes\ in this case is 10.8$\times$10$^{-3}$~h$_{70}^{-3}$~Mpc$^{-3}$, 
which is a factor of $\sim$5 higher than the mean density we observe.  
Taken at face value, this could imply that \laes\ have a duty cycle of $\sim$20\% or 
that they are a younger, less massive subset of this population.  
If we restrict the model galaxies to be younger than the sample's 
median age ($\leq$172~Myr) and less massive than the median stellar mass ($\leq$9$\times$10$^{8}$~$M_{\Sun}$), 
the number density of model \laes\ is 3.5$\times$10$^{-3}$~h$_{70}^{-3}$~Mpc$^{-3}$.  
In either case, we uncover overdensities of greater than a factor of 2 
at a rate of $\approx$0.3\%.  Therefore, within 
the large span of our filter, which will tend to average out inhomogeneities 
and reduce the signal, our imaging survey had only 
a $\approx$0.3\% chance of randomly uncovering such an overdense region 
if \labs\ and overdensities are independent phenomena.  
The space density of large \lya\ nebulae is very uncertain, but at the 
high end is the range quoted for the S00 \lya\ nebulae; 
$\sim$3-400$\times$10$^{-6}$~Mpc$^{-3}$ \citep{sai06}.    
Taking these values (equivalent to a $\sim$17-100\% chance of finding one large \lab\ within 
the peak of the overdensity), the likelihood of a chance coincidence 
between a factor of $>$2 overdensity and 
a large \lya\ nebulae would be $\lesssim$0.05-0.3\%.  
Preliminary results from more recent systematic \lya\ nebulae 
surveys hint that their true space density may 
be orders of magnitude lower (e.g., Prescott et al. 2008, in preparation), 
making the likelihood of chance coincidence vanishingly small.  

In terms of the relevant 
emission mechanisms the \lya\ nebula class appears to be 
a highly heterogeneous mix, and this diversity could in principle derive from 
environmental differences.  The largest \lya\ nebulae ($\gtrsim$100~kpc), 
including the case studied here, often show evidence for obscured AGN and extended 
star formation \citep[e.g.,][Prescott et al. 2008, in preparation]{mat07,basu04} 
and have received the most scrutiny in terms of their properties and environments.  
Presumably, the somewhat smaller ``cooling'' \labs\ \citep[e.g., ][]{nil06,smi07} must 
also reside in dense regions with sufficient gas supply, but 
thorough environmental studies of these sources have, to our knowledge, 
not yet been completed.

\section{Conclusions}

The discovery of a large \lya\ nebula at $z\approx2.7$ via its strong mid-infrared emission 
has provided an unbiased test of the association between 
these rare sources and galaxy overdensities.  Using deep \lya\ imaging 
of the environment surrounding this \lab, we identify 785 \lae\ candidates 
and find evidence for 
a factor of $\sim$3 \lae\ overdensity which spans 20$\times$50~Mpc (comoving).  
This is comparable to what is found in the vicinity of the well-known S00 
\lya\ nebulae.  We rule out a chance coincidence at the $\lesssim$1\% level.  
In conjunction with previous work, these results point conclusively to a strong association 
between the largest \lya\ nebulae and overdense regions of the Universe.

\acknowledgments
We are grateful to Masafumi Yagi 
and Yoshiaki Taniguchi for data reduction advice and assistance.  
We thank Naveen Reddy for the use of data from an ongoing $U_{d}$ imaging program and 
the referee for constructive comments on the manuscript.  
M. P. was supported by an NSF Graduate Research Fellowship.  
Support for this work was provided by NASA (HST-GO10591) 
and the National Science Foundation (Grant No. 0714311).

\clearpage

\end{document}